Title: Sequence-based prediction of function site and protein-ligand interaction by a functionally annotated domain profile database

Running title: Predicting function site and protein-ligand interactions using *fi*DPD

Authors: Dengming Ming[1*], Ming Han[2] and Xiongbo An[2]

Affiliation: [1]College of Biotechnology and Pharmaceutical Engineering, Nanjing Tech University, Nanjing 211816, China PR, [2]Department of Physiology and Biophysics, School of Life Science, Fudan University, Shanghai 200438, China PR

[*]Contact information:
Dengming Ming
Biotech Building Room B1-404
College of Biotechnology and Pharmaceutical Engineering,
Nanjing Tech University,
30 South Puzhu Road
Jiangsu 211816, PR China
Tel: 8625-58139942
Email: dming@njtech.edu.cn

ABASTRACT

**Background**: Identifying protein functional sites (PFSs) and protein-ligand interactions (PLIs) are critically important in understanding the protein function and the involved biochemical reactions. As large amount of unknown proteins are quickly accumulated in this post-genome era, an urgent task arises to predict PFSs and PLIs at residual level. Nowadays many knowledge-based methods have been well developed for prediction of PFSs, however, accurate methods for PLI prediction are still lacking.

**Results**: In this study, we have presented a new method for prediction of PLIs and PFSs based on sequence of the inquiry protein. The key of the method hinges on a function- and interaction-annotated protein domain profile database, called *fi*DPD, which was built from the Structural Classification of Proteins (SCOP) database, using a hidden Markov model program. The method was applied to 13 target proteins from the recent Critical Assessment of Structure Prediction (CASP10/11). Our calculations gave a Matthews correlation coefficient (MCC) value of 0.66 for prediction of PFSs, and an 80% recall in prediction of PLIs.

**Conclusions**: Our method reveals that PLIs are conserved during the evolution of proteins, and they can be reliably predicted from *fi*DPD. *fi*DPD can be used as a complement to existent bioinformatics tools for protein function annotation.

**Background**

Most proteins perform biological functions via interactions with their partners, such as small molecules or ligands, DNA/RNA, and other proteins, forming instantaneous or permanent complex structures. One important fact concerning protein-partner interactions is that only a few pivotal amino acids on protein surface play key roles in determining the interactions. Thus accurately predicting PFSs, and particularly identifying the types of interactions involved in these sites are extermemly important to correct annotation of protein functions, and thus providing valuable information for rational drug design and drug side-effect assessment[1-3]. Up-to-date, the 3D protein-partner complex structures are the main source of knowledge about PFSs and PLIs. In recent years, *in silico* methods have obtained increasing attentions as an alternative strategy for the annotation of protein function. This is due to the two contributions: the accumulation of a large number of 3D protein-complex structures in the publicly accessible Protein Data Bank (PDB)[4], and the fast development in computer technology and computation algorithms. In this post-genome era as piles of uncharacterized protein structure and sequence are quickly accumulated and freely accessible worldwide, a stringent task has arisen in the computational biology community: to develop accurate *in silico* method for prediction of PSFs and involved PLIs[5].

In last decades computational methods had emerged to predict PFSs, which can be roughly grouped into two categories, namely the sequence-based methods and the structure-based methods[6]. Sequence-based methods assume that functionally important residues are conserved through evolution and might be identified as conserved sites based on multiple sequence alignment (MSA) within homologous protein families[7-9]. Sequence-based information such as secondary structure propensity and the likely solvent accessible surface area (SASA) were also calculated and used to improve the function-site prediction[10-13]. Structure-based methods first determine the local or overall structural similarity between the enquiry protein and well-characterized proteins from PDB, and then predict function sites for unknown proteins using PFSs of known ones[14-17]. Typical local structural features include large clefts on protein surface[18, 19], special space-arrangement of catalytic residues[20-22], particular patterns between surface residues[23, 24], etc. Methods that used both structural and sequence information were also developed[25, 26], and when combined with artificial intelligence techniques they might give encouraging prediction results[27-29]. Other methods based on protein dynamics [30-32], protein-ligand docking and conventional molecular dynamics simulations [33-35] were also found successful in PSFs.

As early as the emergence of the first protein-ligand complex structure, researchers had tried to characterize the interactions between proteins and ligands. Very recently, with the rapid accumulation of protein complex structures, structural bioinformatics approaches had been developed to characterize PLIs[36-41], and a bunch of atomic interaction databases were also generated to facilitate PLI studies[42-44]. These data offer new resource for prediction of PFSs and PLIs, and thus provide auxiliary information to facilitate conventional docking simulation and pharmacology researches. However, a systematic usage of these atomic interaction data in PFS prediction and protein function annotation is still lacking.

Recently, we developed a time-efficient method, called fast dynamics perturbation analysis (FDPA), for prediction of ligand binding sites based on structural dynamic information [45-47]. The method was examined with a standard small-molecule docking test-set of GOLD[48], and for 94% of 305 protein complexes FDPA predictions find at least one true binding site. FDPA calculations support the idea that the PFSs tend to be associated with where an external interaction may cause large change in protein conformational distribution. Very recently, we applied FDPA to all available protein domain structures recorded in structural classification of protein database (SCOP)[49], and created a functionally annotated domain profile database (*f*DPD)[50]. In *f*DPD, similar SCOP domain structures are grouped into a profile module on which pivotal sites are marked by mapping PFSs of known proteins in the group. For an inquiry protein, the best-matched modules in *f*DPD are identified based on sequence similarity, using the profile hidden Markov model of HMMER program[51]. Finally, pivotal sites of the modules are collected and ranked as a prediction of PFSs of the inquiry protein. The method had been examined with two standard enzyme datasets: the CATRES-FAM [52] of 140 enzymes with a total of 471 catalytic residues and the Catalytic Site Atlas(CSA-FAM)[53] of 94 enzymes, and was found to obtain 59% recall at precision 18% for the first dataset and 51% recall at 18.5% precision for the second one. In predicting ligand-binding sites of 30 target proteins of CASP9 [54], *f*DPD gave 25 predictions with an average MCC value of 0.56, which ranked between 8[th] and 10[th] of the 33 participating groups[54].

In this paper, we developed a knowledge-based method for prediction of PLIs on the basis of *f*DPD. For this sake, we first generated atomic PLI patterns associated with function sites protein domains in *f*DPD. Then we mapped PLIs of member proteins to pivotal sites of profile modules, thus derived a function-site- and PLI-annotated domain profile database, called *fi*DPD. A similar hidden Markov model of HMMER program was then used to prediction between PFSs and PLIs for inquiry proteins. We applied *fi*DPD method to 10 target proteins of CASP10[55] and CASP11[56], and found it gave correct PLIs for 80% of examined sites. This work is a substantial extension of the original *f*DPD method, and we expect it to be an auxiliary tool to conventional bioinformatics tools for protein function annotations.

## METHODS

Figure 1 shows the flow chart to build *fi*DPD, which is an extension of *f*DPD by taking into account of PLIs. We firstly introduced the *f*DPD as a list of representative profile-modules which are built by sorting out structure-and-sequence similar protein-domains in SCOP databases[57]. Next PFSs and atomic patterns of PLIs were derived from known protein-ligand-complex structures in PDB, and then, after a series of site-to-site mapping, they are used to annotate *f*DPD profile-modules and thus to build the *fi*DPD.

**Domain subgroups extracted from SCOP database are used to build *f*DPD.**

We started with a modified classification of protein-domain structures collected in the SCOP database[57, 58]. In SCOP, a big protein structure is often manually divided into a few smaller parts or domains, according to their spatial arrangement within the protein. A recent version of SCOPe 2.05 was download from http://scop.berkeley.edu/references/ver=2.05, which includes 214547 domain entries extracted from 75226 protein structures in PDB. In SCOP these domain structures are arranged in a hierarchical 7-level system: the Class (*cl*), the Fold (*cf*), the Superfamily (*sf*), the Family (*fa*), the Protein Domain (*dm*), the Species (*sp*), and the PDB code identity (*px*), according to their sequence, function and structure similarity. Particularly, those domains listed in a given domain entry (*dm*) presumably share the same class, same fold, same superfamily and same protein family, but might differ in species and PDB code-entry. Theoretically, PFSs are more likely to be conserved when they share both higher structural and sequential similarity, and this lays the basis for our algorithm of *fi*DPD in prediction of protein function-site and PLIs. Using a profile hidden Markov model of HMMER program, the multiple sequence alignment of all the domains within the same *dm*-entry gives a single profile module representative. In this way, 12527 representative profile modules were created for all the *dm*-entries, which form the basis of *f*DPD and *fi*DPD.

In building *f*DPD, it is important that protein domains within the same *dm*-entry are structurally and sequentially close enough to one another. However, a quick calculation reveals that the $C_\alpha$ root-mean-square-distance (RMSD) can be as large as 12Å or so for some domain structures listed in the same *dm*-entry. This result indicates that many domain structures in the same *dm*-entry of the original SCOPe 2.05 might be quite different, making the profile modules of *f*DPD being less representative of member proteins. To save this situation, we divided the domains within a *dm*-entry into a few smaller groups or subgroups so that selected domains within the same subgroup have mutual $C_\alpha$-RMSD < 7Å and mutual sequence-similarity > 10 (a score value calculated by the multiple sequence alignment program of CLUSTRALW[59]). Thus derived subgroups are then replace the "*dm*"-entry as the basic unit of *f*DPD. *f*DPD contains 16559 subgroups, which is 32% more than the original SCOP *dm*-entries with about 12 member structures on average in each subgroup.

**_f_DPD is a list of profile modules which are protein-domain representative.**

In *f*DPD, protein sequences of all domains in a subgroup were extracted and aligned using the multiple sequence alignment (MSA) program MUSCLE[60], from which a profile module was then built using the *hmmbuild* module of the HMMER program(http://hmmer.org/[61]). A profile module is a sequence of hypothetical amino acids, which is, instead of conventional amino acid, probably a mixture of certain amino acids according to the MSA of the subgroup. For each individual position in a profile module, we defined a conservation value $C$, according to the MSA. We assigned $C$ value as 0, 1, 3, 4 for a position being non-conservative, little-conservative, conservative and highly conservative, as indicated respectively by gap, symbol "+", small letter or capital letter in the MUSCLE alignment. We also defined an overall volume value $N$ for a profile module as the number of protein domains listed in the subgroup: a larger $N$ value usually means more knowledge available for that subgroup and thus greater confidence on the annotation.

**_f_DPD profile modules are annotated with function-sites of member proteins.**

A scoring function $S$ was assigned to each position of an *f*DPD profile module to mark its propensity of being a function site. To do so, we first mapped known function-sites of member proteins within the same subgroup to the profile module, according to the MSA(see Figure 2). Function sites of member proteins were collected from the SITE sections of the corresponding PDB file. Of the 202705 protein domains listed in SCOPe 132725 domain structures have a total of 1878004 function-sites annotated in PDB SITE records. Then, for simplicity, we assigned $S$ as the total hit-number that a profile module position receives based on the MSA. Thus, the larger a position has $S$ value the more likely it might be a hypothetical function site for the profile-module. In this way, the profile modules are annotated with known PFSs, thus called the *function-site* annotated domain profile database or *f*DPD. Previously, alternative function-site annotations for profile-modules were also built by using different "known" PFSs as derived from FDPA calculations[50].

**_fi_DPD was derived from _f_DPD by adding PLI annotations to the profile modules.**

Obviously, the above-mentioned $S$-value is heavily depended on the ways with which the "known" PFSs are determined. In this work, $S$-values are determined by only using PDB SITE information, which, in most case, are manually prepared ligand-binding sites. Other types of biologically-interested function-site data, such as enzyme active site[53], phosphorylation sites[62], etc., might also be used in the annotation. Here, considering the importance of PLIs in determining protein function, we added PLI annotations to the profile-modules of *f*DPD, so as to build the *function-site* and interaction annotated domain profile database or *fi*DPD.

To annotate the profile modules with PLIs, atomic interaction patterns between protein and ligand were firstly determined based on their 3D protein-ligand complex structures. Specifically, atomic 3D coordinates of amino acids listed in PDB SITE sections and those of ligand molecules were firstly filtered out from PDB files, then a series of atomic distance ($d$) were calculated between PFSs ($A_{Site}$) and ligands ($A_{Ligand}$), and finally a few types of bonding and non-bonding interactions for each $A_{Site}$ were determined based on the pair-wise distances and the biochemical properties of involved amino acids.

***Covalent bond and coordinate bond*** Usually non-bonded forces dominate interactions between a ligand and its target protein, however irreversible covalent bonds are also found in PLIs when tight-and-steady connection between ligand and receptor is essential to the biological function, such as in the rhodopsin system[63]. A covalent bond is built if the distance between a non-hydrogen atom from a function site and a non-hydrogen atom from ligand satisfies $d < R(A_{Site}) + R(A_{Ligand}) + 0.5\text{Å}$, where $R(A)$ is the radius of atom A. For metal-ion ligands, this also defines coordinate bonds between metal ions and PFSs. Usually, in coordinate bonds the shared electrons are presented in atoms with higher electronegativity in a function site. We denoted NCOV as the total number of covalent bonds involving atoms in the function site, and NCOO as the total number of coordinate bonds involving atoms in that site.

***H-Bond*** Almost all PLIs occur in aqueous environment where water molecules play critical role. As a result, hydrogen bonds might be consistently established and destroyed until certain stable protein-ligand configuration is achieved. Here we have calculated hydrogen bonds within the protein-ligand complex using the program HBPLUS[64]. The program determines H-Bond donor (D) and acceptor (A) atom pairs based on a non-hydrogen-atom configuration, using a maximum H–A distance of 2.5Å, maximum D–A distance of 3.9Å, minimum D–H–A angle of 90° and minimum H–A–AA angle 90°, where H is the theoretical Hydrogen atom, AA is the atom of function sites in H-bond acceptor. In this way, we defined NHBA and NHBD as the total number of H-bond acceptors and H-bond donors associated with atoms in a given function site.

*Electrostatic interactions*   Electrostatic force plays important roles in many PLIs, and might be the main driving forces to initiate catalytic reactions, to guide the recognition between protein and ligand, and so on[65-67]. However, accurately determining atomic charges in bio-structure is a very challenging task since it is highly sensitive to the surrounding environment. Here, for simplicity, we identified electrostatic interactions simply by examining the charging status of contact atoms in PLIs. Specifically, we first selected positively charged nitrogen (N) atoms of function sites of Arg, His, and Lys, and then determined an electrostatic interaction if there exists a neighboring (< 4.5 Å) oxygen atom in ligand which is not a part of a cyclized structure. An electrostatic interaction was also built when a negatively charged oxygen (O) atom from Asp and Glu residues is found near a ligand nitrogen atom. We used NELE as the total number of electrostatic interactions involving atoms in a give function site.

*π-stacking interaction*   π-stacking interactions play critical role in orientating ligands inside binding pockets. We first identified aromatic side chains of Trp, Phe, Tyr and His of protein function-sites and carbon-dominant cyclized structures of ligands. Usually, aromatic rings form an effective π-stacking interaction when they get close enough (4.5~7Å) and have either parallel or perpendicular orientation[68, 69]. Here for simplicity, we defined a π-stacking interaction if we can find three or more distinct heavy-atom-pairs between atoms from aromatic ring of given function site and those from ligand carbon-ring structures. We defined the total number of π-stacking interactions involving a given function site as NPI.

***Van der Waals interaction***    A Van der Waals interaction is built when the distance $d$ between a non-hydrogen atom of protein function-site and a non-hydrogen atom of ligands satisfies,

$$d < \text{vdW}(A_{\text{Site}}) + \text{vdW}(A_{\text{Ligand}}) + 0.5\text{Å},$$

where $\text{vdW}(A)$ is the van der Walls radius of atom A, and no covalent bond, coordination bond, hydrogen bond, electrostatic force and π-stacking interaction are found between them. The atomic van der Waals radii are taken from the CHARMM22 force-field[70]. Each function site was assigned a NVDW value as the total number of Van der Waals interactions involving atoms of this site.

Taken together, we characterized a PLI between a PFS and the ligands with a 7-dimensional interaction vector **v** = (NCOV, NCOO, NHBA, NHBD, NPI, NELE, NVDW). Interaction vectors of all member protein were added up in different pivotal sites of the profile-module according to the MSA of the studied subroup. As a result, each *f*DPD profile-module was annotated with interaction vectors **V**'s on hypothetical function-sites, thus forming the *fi*DPD.

***fi*DPD predicts functional sites and PLIs by a hidden Markov model.**

*fi*DPD is essentially a list of profile module entries that are annotated with domain function sites and PLIs. In *fi*DPD, a two-steps are required to predict hypothetical functional-sites and involved PLIs for a given enquiry protein: 1) identifying profile modules in *fi*DPD that match the inquiry sequence best, and 2) interpreting pivotal function sites and associated PLIs of the matched profile modules as a prediction of PFSs and PLIs for inquiry protein, based on certain statistical evaluations.

In the first step, *fi*DPD scans the inquiry sequence against all its module-entries, using the SCAN module of the HMMER program[71]. The scan usually gives a couple of profile modules within an alignment E-value cutoff no greater than $1\times10^{-5}$. Each alignment (indexed by superscript *j* in Equation (1)) is assigned a scoring function ***E*** as the negative logarithm of the E-value score. Due to the limited volume of known protein sequences contained in *fi*DPD, there are cases in which HMMER SCAN cannot find any match for inquiry protein, and for these cases *fi*DPD simply gives a notice of "no-hit". In step 2), we defined a scoring function $F_i$ for the *i*th residue of the inquiry protein as its propensity to be a function-site:

$$F_i = \sum_j S_{i'}^j C_{i'}^j N^j E^j \tag{1}$$

where the summation runs over all the alignments *j*, and $i'$ stands for the position of profile module that matches the *i*th residue of inquiry protein. Residues with high-valued ***F***-scoring function will be predicted as hypothetical functional sites.

One way to determine high-***F***-valued sites for inquiry protein is simply to choose a certain number (*n*) of top-valued residues, called *n*-top selection. This method had been used for enzyme catalytic site prediction[50], since experimentally determined enzyme active sites have relatively fixed number as revealed by the dataset Catalytic Site Atlas (CSA)[53]. Another method to select top-valued residues uses a cutoff percentage that was proved to be efficient in a previous ligand-binding site prediction[72, 73]. In this method, we firstly filtered out those low-valued noise-like residues whose ***F***-scores are smaller than a cut-off percentage $M\%$ of the maximum ***F***–value $F_{max}$, and then, for the rest residues, the top $T\%$ residues are predicted as hypothetical function-sites of the inquiry protein. Usually, in this selection strategy, a larger-sized protein tends to have more hypothetical function sites. We used this selection strategy to predict PFSs in the rest of this paper.

To predict PLIs, we defined an interaction scoring-vector $\boldsymbol{I}_i$ for the $i$th residue of the inquiry protein, following Equation (1):

$$\boldsymbol{I}_i = \sum_j \boldsymbol{V}^j_{i'} C^j_{i'} N^j E^j \qquad (2)$$

where $\boldsymbol{V}^j_{i'}$ is interaction vector of the $i'$th position in the profile module $j$ that corresponds to the $i$th residue of the inquiry sequence. For each prediction function site, $fi$DPD will determine an associate PLI vector according to Equation (2).

*Validation datasets*

The original $f$DPD had been examined for PFS prediction using a few types of datasets, including two manually cultivated enzyme-catalytic site datasets of 140-enzyme CATRES-FAM[52] and 94-enzyme Catalytic Site Atlas (CSA-FAM)[53] and a 30-membered small-molecular binding protein targets from CSAP9[74]. Here we examined $fi$DPD by calculating PLIs of protein targets listed in CASP10[75] and in CASP11[56] whose ligand-binding complex structures had been solved.

*Validation method*

The conventional prediction precision and recall were used to evaluate the performance of our method: $\text{Precision} = \text{TP}/(\text{TP} + \text{FP})$ and $\text{Recall} = \text{TP}/(\text{TP} + \text{FN})$, where the true positive (TP) is the predicted residues listed as functional sites in the dataset, the false positive (FP) is the predicted not listed in the dataset, and the false negative (FN) is the functional sites listed in the dataset but missing by the method; another relevant quantity is the true negative (TN) that stands for the correctly predicted non-binding/non-functional site residues. In our calculations the statistics did not take account of the "no-hit" predictions. The overall precision is the sum of all the TPs divided by the total number of predicted residues, and the overall recall is the sum of all the TPs divided by the total number of listed functional sites in the dataset. The precision-recall curve was found slightly dependent on the cutoff percentage M% and T% in the selection method. The Matthews correlation coefficient[76] (MCC) was used to assess the ligand-binding residue predictions of the CASP10 target proteins[54], which is defined as follows:

$$\text{MCC} = \frac{\text{TP} \times \text{TN} - \text{FP} \times \text{FN}}{\sqrt{(\text{TP}+\text{FP}) \cdot (\text{TP}+\text{FN}) \cdot (\text{TN}+\text{FP}) \cdot (\text{TN}+\text{FN})}}.$$

The predicted PLIs were compared with those directly derived from 3D protein-ligand complex structures, and a precision and recall value was obtained to qualify PLI predictions.

**RESULTS AND DISCUSSIONS**

**The mimivirus sulfhydryl oxidase R596**

The 292aa mimivirus sulfhydryl oxidase R596 is target T0737 of CASP10 whose structure was later determined at 2.21Å (PDB entry 3TD7, see Figure 3[77]). The protein comprises of two all alpha-helix domains, the N-terminal sulfhydryl oxidase domain (Erv domain) and the C-terminal ORFan domain. The enzyme minivirus R596 has an EC number of EC1.8.3.2, catalyzing the formation of disulfide bonds through an oxidation reaction with the help of a cofactor of flavin adenine dinucleotide (FAD). FAD is found tightly bonded to 22 residues in the catalytic pocket in Erv domain[55], playing an important role in transferring electrons from a 10Å-distance shuttle disulfide in the flexible inter-domain loop to the active-site disulfide close to FAD in Erv domain[77]. In the prediction, *fi*DPD scanned the T0737 sequence against the database and found 4 profile module entries, all from the Apolipoprotein family with a structure of four-helical up-and-down bundle. The 4 entries include an automated-match-domain profile built from 10 sequences from *Arabidopsis thaliana*, a second automated-match-domain profile built from 4 sequences from *Rattus norvegicus*, an augmenter of liver regeneration domain profile built from 13 sequences from *Rattus norvegicus*, and a thiol-oxidase Erv2p domain profile built from 6 sequences from *Saccharomyces cerevisiae*. The scanning E-value ranges from $2\times10^{-8}$ to $1\times10^{-19}$, indicating that the inquiry sequence only has moderate similarity with the annotated sequences in the database. A total of 56 annotated pivotal sites in the 4 *fi*DPD profile modules were then collected and sorted according to their function-site scoring functions. When mapping to the inquiry sequence, 13 function sites were then automatically identified, which leads to a 100% prediction precision and 59% recall. We also examined those function sites that *fi*DPD failed to identify and found that they actually locate in a different C-terminal domain other than the four-helical up-and-down bundle domain.

To examine the PLI prediction, we first collected interaction scoring-vectors associated pivotal sites in the four profile modules according Equation (2), and then compared with those directly determined from the protein-ligand complex structure recorded in PDB entry 3TD7(Table 1). Figure 3 demonstrates key interactions predicted by Equation (2) and those not found by the prediction. *fi*DPD correctly predicted all the π-stacking interaction involving Trp45, His49, Tyr114, His117, indicating that π-π interactions play critically important role in ligand binding. The prediction also found significant π-stacking interactions on pivotal sites of Leu78 and Lys123, however these π-π interaction predictions were ignored in a post-treatment simply because of lacking of aromatic side-chains in these residues. *fi*DPD also finds the correct electrostatic interactions on sites of His117 and Lys123. It identified large probability of electrostatic interactions on sites of Thr42 and Val126, however these interactions were ignored in a post-treatment since the involved residues are not chargeable in the conventional conditions. Taken together, the overall PLI predictions associated with the identified function sites are about 80%.

**CASP10 Targets and CASP11 Targets**

We applied *fi*DPD to protein targets listed in CASP10 and CASP11, of which 13 targets had been solved to have explicit ligands bound[55]. Table 2 listed all the predictions, of which *fi*DPD gave no-hit for 3 target proteins. For the rest 10 predictions, *fi*DPD gave an overall precision of 87.7% and the overall recall of 59.9%, using a scale-selection with T% of 45% and M% of 35%. The averaged MCC of 0.66 which ranked between 4$^{th}$ and 5$^{th}$ of the 17 participating groups[54]. Considering the ligand-binding types, we found that *fi*DPD gave better function-site predictions for metal binding sites with an averaged MCC value of 0.68 while it is 0.46 for non-metal binding sites prediction, indicting that PFSs are more conservative with respect to either space arrangement or sequence location in metal binding.

We examined the performance of *fi*DPD prediction of PLIs in these target proteins, by determining the overlap between the predicted PLIs and those calculated based on solved protein-ligand complex structures. Table 3 compared the predicted PLIs on function sites with the experiment PLIs. In most cases, *fi*DPD can correctly identified 80% or more of PLIs on function sites.

**Conclusions**

In this paper we presented a function-site- and PLI-annotated domain-profile database or *fi*DPD, from which we developed a sequence-based method for prediction of PFSs and PLIs. Our method is based on the belief that proteins that share similar structure and sequence tend to have similar functional sites located on the same positions on protein surface. A profile module entry in *fi*DPD is a representative of a bunch of protein domains that share high sequence-and-structure similarity as recorded in the SCOP database. For an inquiry protein, *fi*DPD method identifies profile modules in the database using the hidden Markov model algorithm of HMMER program, and then maps the annotated pivotal sites and associated PLIs of the profile modules to residues in the inquiry protein as a prediction.

In earlier studies, *f*DPD method was examined in the prediction of catalytic sites of a standard dataset of 140-enzyme CATRES-FAM[52], resulted in an enzyme active-site prediction of 59% recall at a precision of 18.3%. For ligand-binding site prediction of target proteins in CASP9, *f*DPD obtained an averaged MCC of 0.56, ranking between $8^{th}$ and $10^{th}$ of the 33 participating groups[54]. *fi*DPD is identical to *f*DPD in prediction of PFSs, however, since its database is annotated with PLIs, *fi*DPD can give new prediction of PLIs on predicted function-sites. Here, *fi*DPD was applied to predict function sites of 10 target proteins in CASP10 and CASP 11 that have been solved in a ligand-bound state, and an averaged MCC of 0.66 was achieved. Further more, we compared the predicted PLIs with those directly calculated based on 3D protein-ligand-complex structures, and found that the predicted PLIs correctly overlap 80% of the true PLIs. Our calculations indicate that the PLIs are well-conserved biochemical properties during the protein evolution, and it is possible to assign accurate PLIs on predicted PFSs using annotated database.

There are a few ways to improve the performance of *fi*DPD predictions. First, adding new annotations to profile modules of *fi*DPD, which enable *fi*DPD to make new type of predictions. For example, mapping enzyme catalytic site (CSA), Zinc-binding sites, RNA-binding sites to profile modules should largely improve *fi*DPD prediction of catalytic sites, Zinc-binding sites, RNA-binding sites. As a knowledge-based method, the utility and efficiency of *fi*DPD prediction suffers from the sampling limitation of annotations of known proteins. However, this sampling problem might be partially solved with the efforts of large scaled protein sequencing and the worldwide structure genomics projects. There still remains a challenge for protein function-annotation researcher to predict more and more detailed information about protein functioning, and to meet this task, it obviously needs the extensive cooperation between accurate theoretical modeling calculations and a large amount of experimental validations. *fi*DPD provides an example that atomic PLIs might be predicted from protein sequence, using a new type of function-site and PLI-annotated database.

**Acknowlegement**

We appreciate Prof. Rupu Zhao for his careful reading the manuscript and helpful comments.

**Declarations**

**Funding**

This work was supported, in part, by the National Natural Science Foundation of China (Grant No. 31270759).

**Availability of data and materials**

The *fi*DPD database is available from the corresponding author on reasonable request.

**Authors' contributions**

DM designed *fi*DPD algorithm. DM and MH wrote code of the *fi*DPD program. MH performed the computational experiments. DM wrote the paper. All authors read and approved the final manuscript.

**Competing interests**

The authors declare that they have no competing interests.

**Consent for publication**

Not applicable.

**Ethics approval and consent to participate**

Not applicable.

REFERNECE

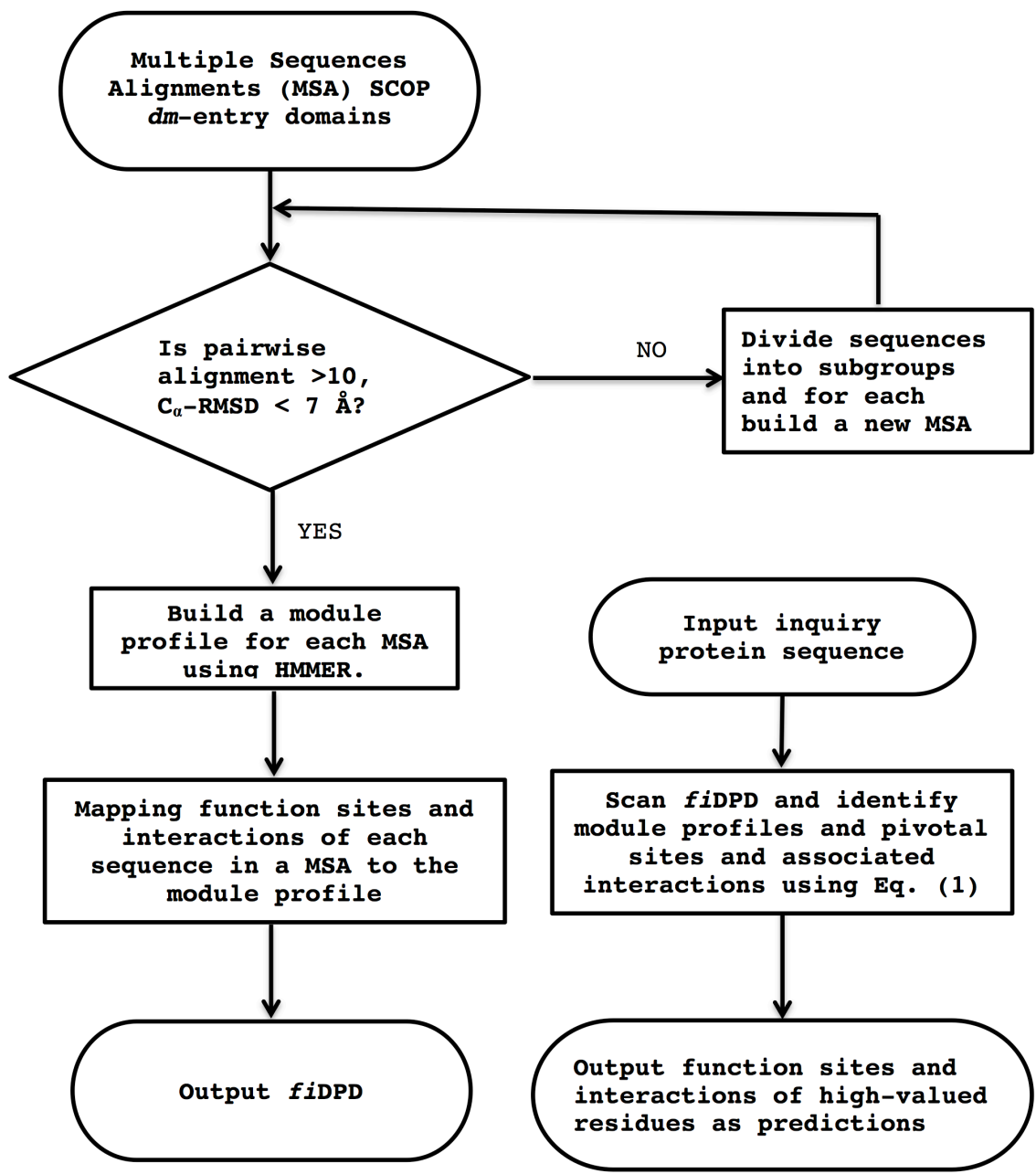

Figure 1. Flow-chart for building the function-site- and interaction-annotated domain profile database (*fi*DPD) and for predicting protein function-sites and PLIs using *fi*DSPD.

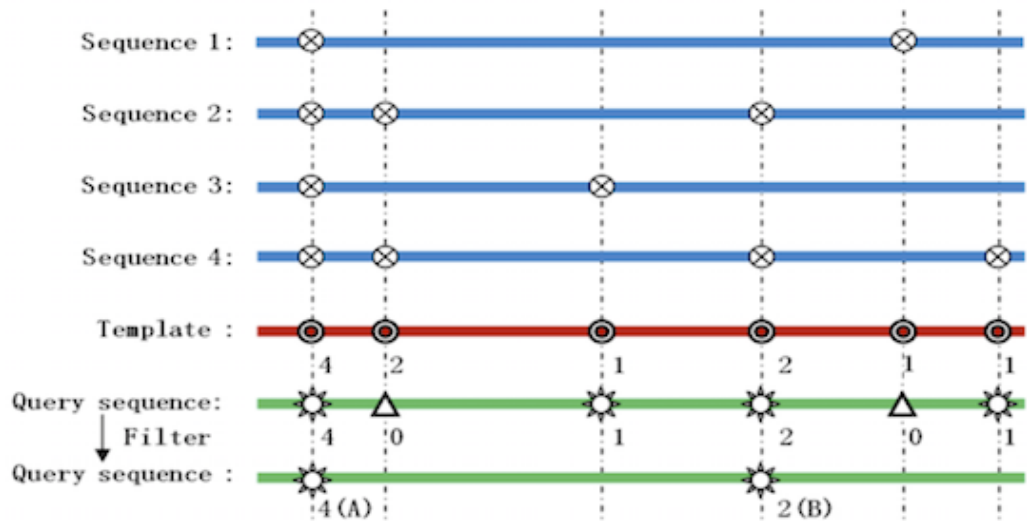

Figure 2. Mapping known PFSs to profile module, ⊗: known PFSs of domain structures, ⊙: pivotal PFSs in a profile module with the number indicating a weight factor, ✶: PFSs mapped into the inquiry protein sequence from profile module pivotal sites, which, after a filtering, is reduced to two points (A and B) as a final prediction output, △: non-conservative pivotal sites mapped into the inquiry protein, which will be ignored due to the low conservation value.

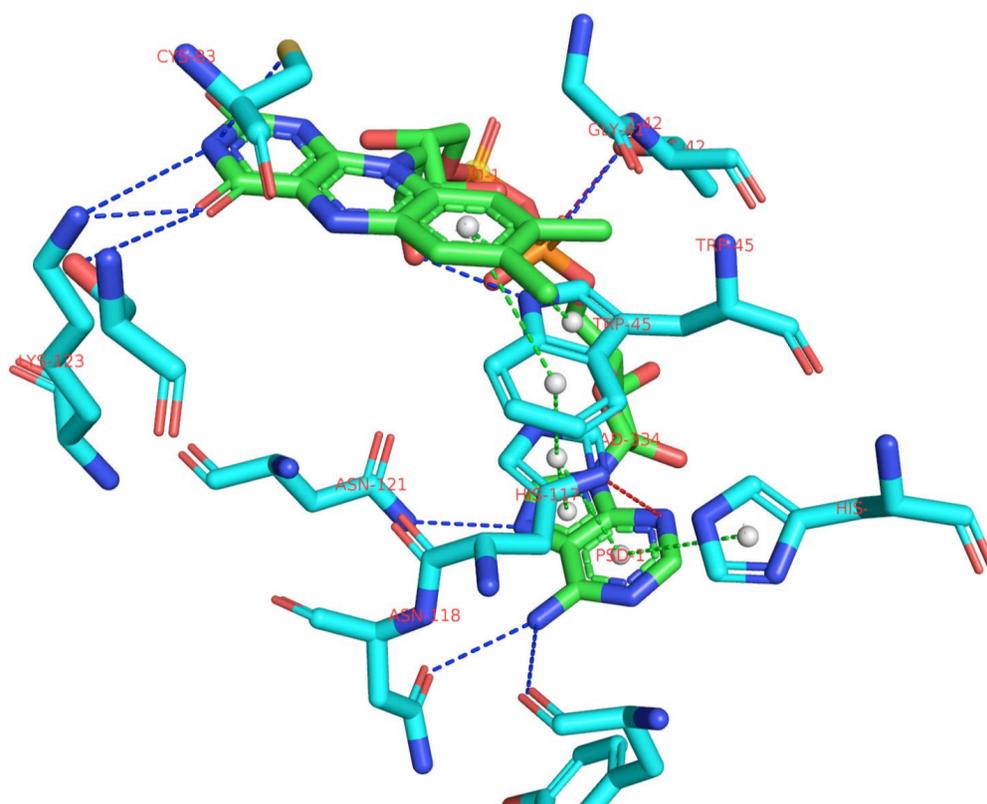

Figure 3. Mapping the protein-ligand interactions predicted for *the mimivirus sulfhydryl oxidase R596*, target T0737, PDB code 3TD7. Dash lines represent PLIs, they are colored as following: blue for electrostatic interactions, green for π-stacking interactions, gray for van der Waals interactions, and red for interaction not found by *fi*DPD.

Table 1. The prediction of protein-ligand interactions on PFSs of T0737[†].

| Target | Site | AA | COV | COO | ELE | HBD | HBA | π-π |
|---|---|---|---|---|---|---|---|---|
| T0737 | 41 | G | 0 | 0 | 0 | 0 | 0 | 0 |
| | 42 | T | 0 | 0 | +/0 | T | 0 | 0 |
| | 45 | W | 0 | 0 | 0 | T | 0 | T |
| | 49 | H | 0 | 0 | 0 | 0 | + | T |
| | 78 | L | 0 | 0 | 0 | 0 | 0 | 0 |
| | 83 | C | 0 | 0 | 0 | + | T | 0 |
| | 114 | Y | 0 | 0 | 0 | 0 | T | T |
| | 117 | H | 0 | 0 | T | + | - | T |
| | 118 | N | 0 | 0 | 0 | + | T | 0 |
| | 120 | V | 0 | 0 | 0 | 0 | 0 | 0 |
| | 121 | N | 0 | 0 | 0 | 0 | + | 0 |
| | 123 | K | 0 | 0 | T | T | + | +/0 |

†AA stands for amino acid, COV for covalent bond, COO for coordinate bond, ELE for electrostatic interaction, HBD for H-bond donor, HBA for H-bond acceptor, π-π for π-stacking interactions. "0" indicates the corresponding interaction is not present in protein-ligand complex structure and *fi*DPD calculation also showed no such type PLIs on the site.

Table 2. Predictions of CASP10 targets proteins[†].

| Target | PDB | Ligand | Type | Sites | Prediction | TP | Precision | Recall | MCC |
|---|---|---|---|---|---|---|---|---|---|
| T0652 | 4HG0 | AMP | Non-metal | 13 | 31 | 12 | 0.57 | 0.92 | 0.71 |
| T0657 | 2LUL | ZN | Metal | 4 | 14 | 4 | 0.29 | 1 | 0.52 |
| T0659 | 4ESN | ZN | Metal | 3 | No-hit | | | | |
| T0675 | 2LV2 | ZN | Metal | 8 | 5 | 5 | 1 | 0.62 | 0.77 |
| T0686 | 4HQL | MG | Metal | 3 | 2 | 2 | 1 | 0.67 | 0.81 |
| T0696 | 4RT5 | NA | Metal | 3 | 2 | 2 | 1 | 0.67 | 0.81 |
| T0697 | 4RIT | LLP | Non-metal | 14 | 11 | 10 | 0.91 | 0.71 | 0.8 |
| T0706 | 4RCK | MG | Metal | 6 | 3 | 3 | 1 | 0.5 | 0.7 |
| T0720 | 4IC1 | MN/SF4 | Metal | 14 | No-hit | | | | |
| T0721 | 4FK1 | FAD | Non-metal | 31 | 3 | 3 | 1 | 0.1 | 0.3 |
| T0726 | 4FGM | ZN | Metal | 3 | No-hit | | | | |
| T0737 | 3TD7 | FAD | Non-metal | 22 | 13 | 13 | 1 | 0.59 | 0.76 |
| T0744 | 2YMV | FNR | Non-metal | 19 | 4 | 4 | 1 | 0.21 | 0.45 |

Average[*]

† "Site" stands for number of binding sites found the protein-ligand complex structures.

Table 3. PLI predictions of CASP10/11 targets proteins[†].

| Target | Interactions | Correct Prediction | Recall |
|--------|--------------|--------------------|--------|
| T0652 | 60 | 36 | **60%** |
| T0657 | 24 | 23 | **95.80%** |
| T0675 | 30 | 28 | **93.30%** |
| T0686 | 18 | 17 | **94.40%** |
| T0696 | 18 | 15 | **83.30%** |
| T0697 | 104 | 72 | **69.20%** |
| T0706 | 24 | 21 | **87.50%** |
| T0720 | 78 | 58 | **74.40%** |
| T0721 | 60 | 50 | **83.30%** |
| T0737 | 72 | 63 | **87.50%** |
| T0744 | 42 | 37 | **88.10%** |
| T0762 | 42 | 35 | **83.30%** |
| T0764 | 60 | 52 | **86.70%** |
| T0770 | 18 | 14 | **77.80%** |
| T0784 | 18 | 18 | **100%** |
| T0854 | 24 | 20 | **83.30%** |

† Target 762 to 854 were taken from CASP11 whose protein-ligand interactions were well characterized in the crystal structures.